\documentclass[a4paper,UKenglish, autoref, thm-restate, 11pt]{article}

\usepackage{amsmath}
\usepackage{amsthm}
\usepackage{mathtools}
\usepackage{amsfonts}

\usepackage[hidelinks]{hyperref}
\usepackage[capitalize, nameinlink]{cleveref}

\usepackage[margin=3cm]{geometry}

\usepackage{tikz-cd}
\usepackage{faktor}

\newtheorem{theorem}{Theorem}[section]

\theoremstyle{definition}
\newtheorem{point}[theorem]{}
\newtheorem{lemma}[theorem]{Lemma}
\newtheorem{definition}[theorem]{Definition}

\newtheorem{remark}[theorem]{Remark}

\counterwithin{subsection}{section}

\usepackage[ruled,vlined]{algorithm2e}
\usepackage{algpseudocode}

\usepackage{float}
\SetKwInput{KwIn}{I}
\SetKwInput{KwOut}{O}

\DeclareMathOperator{\rnk}{rk}

\newcommand{\lowp}[1]{low\left(#1\right)}
\newcommand{\leftp}[1]{left\left(#1\right)}

\title{Notes on pivot pairings}

\author{Barbara Giunti\\ {\small Universit{\`a} degli Studi di Modena e Reggio Emilia, Italy}\\ {\small Institute of Geometry, TU Graz, Austria}\\ {\small bgiunti@tugraz.at}}

\date{}

\begin{document}

\maketitle

\begin{abstract}
	We present a row reduction algorithm to compute the barcode decomposition of persistence modules.
	This algorithm dualises the standard persistence algorithm and clarifies the symmetry between clear and compress optimisations.
\end{abstract}

\section*{Introduction}

Persistent homology is a tool of Topological Data Analysis (TDA) whose applications range widely from biology to urban planning to neuroscience (see \cite{database} for an extended list of applications).
Persistent homology summarises a dataset's information in the form of barcode \cite{barcode_carlssonetal, barcode_ghrist}. 
The efficient computation of such barcodes is one of the main
algorithmic problems in TDA \cite{roadmap}.

We provide an algorithm (Algorithm \ref{algorithm_row_pivot_red}) for the barcode decomposition based on row pivot pairing, which is the dual of the column pivot pairing presented in \cite{vines, edels_harer}. 
The algorithm reduces the boundary matrix of a given filtered simplicial complex proceeding by rows and performing row additions, allowing the full exploitation of the compress optimisation \cite{clearcompress}.

A non-exhaustive list of algorithms to decompose persistence modules into interval modules includes the so-called standard algorithm \cite{edels_harer}, the chunk algorithm \cite{clearcompress}, the twist algorithm \cite{twist}, the pHrow algorithm \cite{dualities}, and Ripser \cite{ripser_paper}. 
The implementation of the first four can be found in \cite{PHAT} (cf. \cite{phat_paper}), and that of Ripser in \cite{ripser_software}.
All of them take as input a filtered simplicial complex and employ column operations.
Moreover, in a recent and independent work \cite{Edel-Olsb}, a dualisation of the standard algorithm is presented, using row operations.
In addition to these algorithms, in \cite{abcw} it is shown that the barcode decomposition can also be achieved via the decomposition of filtered chain complexes, whose reduction is performed by row operations.
However, the reduction in \cite{abcw} is different from the standard one, which is the focus of this work. 
Thus, we will not further study the algorithm presented in \cite{abcw}.

As we mention, the idea of proceeding by row is not new. 
Here, we use the straightforward idea of row pivots to clarify the duality between clear and compress optimisations in the computation of persistent homology and cohomology.
In \cite{ripser_paper, phat_paper, dualities}, it is shown that, for Vietoris-Rips complexes, the column reduction, coupled with the clear optimisation and applied to the coboundary matrix, provides a significant speed-up in the computation of the barcode. 
This improvement is not mirrored when the same reduction is coupled with the compress optimisation and applied to the boundary matrix \cite{ripser_paper, clearcompress, dualities}.
Here, we show that the reason for this asymmetry is that the second procedure is not the true dual of the first one: to obtain the dual, and thus the same number of operations, it is necessary to reduce the boundary matrix via row operations instead of via column operations.

\section{Preliminaries}\label{section_preliminaries}

Throughout the work, the symbol $K=\{\sigma_1, \dots, \sigma_m\}$ denote a simplicial complex of dimension $d$, such that for each $i\leq m$, $K_i\coloneqq \{\sigma_1, \dots, \sigma_i\}$ is again a simplicial complex. 
The chain $\emptyset=K_0\subset K_1\subset\dots \subset K_m=K$, denoted by the symbol $FK$ throughout the work, is called a {\bf filtration} of $K$.

Given a simplex $\sigma$ of dimension $h$, its {\bf boundary} is the set of the faces of $\sigma$ in dimension $h-1$. 
If $h=0$, the boundary is empty. 
Otherwise, it consists of exactly $h+1$ simplices. 
The {\bf boundary matrix} $\partial$ of $FK$, is a $(m\times m)$-matrix with coefficients in $\mathbb{F}_2$ where the $j$-th column represents the boundary of $\sigma_j$, i.e. $\partial[i,j]= 1$ if and only if $\sigma_i$ is in the boundary of $\sigma_j$. 
Note that, since $\sigma_i$ is in the complex before $\sigma_j$ is added, $\partial$ is an upper-triangular matrix.

For $0\leq i \leq m$ and $0\leq p\leq d$, we denote the $p$-th homology of $K_i$ over the field $\mathbb{F}_{2}$ by $H_p(K_i)$. 
The inclusions $K_i\hookrightarrow K_j$, for all $i\leq j$, induce maps on the homology vector spaces $H_p(K_i)\to H_p(K_j)$ for all $0\leq p\leq d$.
This collection of vector spaces and maps forms a diagram called a {\bf persistence module} (Fig.~1(a)) \cite{chazal-oudot}.
\begin{figure}[H]\label{fig_pers_int_mod}
	\centering
	\begin{tikzcd}[row sep=0.1cm]
	(a)
	& 0 \ar[r]
	& V_{1} \ar[r]
	& V_{2} \ar[r]
	& \cdots \ar[r]
	& V_{m-1} \ar[r]
	& V_{m}
	\\
	(b)
	& 0 \ar[r]
	& \mathbb{F}_2 \ar[r, "\mathbf{1}"]
	& \mathbb{F}_2 \ar[r, "\mathbf{1}"]
	& \cdots \ar[r, "\mathbf{1}"]
	& \mathbb{F}_2 \ar[r]
	& 0
	\\
	& \scriptstyle\cdots
	& \scriptstyle a
	& \scriptstyle a+1
	& \scriptstyle \cdots
	& \scriptstyle b
	& \scriptstyle b+1
	\end{tikzcd}
	\caption{(a) A persistence module; (b) An interval module with interval $[a,b]$.}
\end{figure}

Since we consider only simplicial complexes with finitely many simplices, the vectors spaces are finite-dimensional.  
In this case, a persistence module admits a nice decomposition into interval modules, which consist of copies of $\mathbb{F}_2$ connected by the identity morphism for the indices inside an interval, and zero otherwise (Fig.~1(b)) \cite{crawley-boevey}.
The collection of intervals in the decomposition of a persistence module is called a {\bf barcode} \cite{barcode_carlssonetal, barcode_ghrist}, and it is an invariant of isomorphism type.
In \cite[Sec.~VII]{edels_harer}, the standard algorithm to retrieve the barcode of a filtered simplicial complex is described. 
This algorithm retrieves the barcode by studying the lowest elements in the columns of the boundary matrix, whose indices form the so-called (column) pivots (see Definition~\ref{def_pivot_column}).
Namely, the algorithm performs left-to-right column additions on columns with the same pivot until no two columns share the same pivot.

\section{Row vs column pivot pairing}\label{section_eq_relations}

The (column) pivot pairing in a reduced boundary matrix $\partial$ provides the lifespan intervals of the homological features of a persistence module \cite{edels_harer}.
Usually, the reduction is performed using only one type of elementary column operation: adding a column to a later column.
Here, we prove that also a reduction performed using only one type of elementary row operation, namely adding a row to a previous row, achieves the same pairing (cf.~\cite{Edel-Olsb}).
The reason why other types of elementary row (column) operations are not allowed is that they do not preserve the order of the generators, and thus cannot maintain the pairing.
\medskip

Let $FK$ be a filtered simplicial complex, as described in \cref{section_preliminaries}, with boundary matrix $\partial$.
For the $i$-th row of $\partial$, let $\leftp{i}$ denote the column index of the leftmost element of such row.
If row $i$ is zero, set $\leftp{i}=0$.

\begin{definition}\label{def_pivot_row}
	A matrix $R$ is called {\bf row reduced} if $\leftp{i}\neq \leftp{i'}$ for all non-zero rows $i\neq i'$.
	An index $j$ is called a {\bf row pivot} (of $R$) if there exists a row $i$ of $R$ such that $j=\leftp{i}$. 
\end{definition}

We recall the some standard notions from \cite{edels_harer}.
The symbol $\lowp{j}$ denotes the index row of the lowest element of column $j$.
If column $j$ is trivial, then $\lowp{j}=0$.

\begin{definition}\label{def_pivot_column}
	A matrix $C$ is called {\bf column reduced} if $\lowp{j}\neq \lowp{j'}$ for all non-zero columns $j\neq j'$.
	An index $i$ is called a {\bf column pivot} (of $C$) if there exists a column $j$ of $C$ such that $j=\lowp{i}$.
\end{definition}

Algorithm \ref{algorithm_row_pivot_red}, called {\bf row pivot reduction}, takes as input the boundary matrix $\partial$ of a filtered simplicial complex $FK$ and reduces it by row operations. 
This algorithm is one of the possible methods to achieve a row reduced matrix.
Indeed, several different row reduced matrices can be obtained by the same boundary matrix $\partial$. 
This follows from the fact that, to the right of each row pivot, there can be several non-zero elements that do not affect the row pivots.

Let $m$ be the number of rows of $\partial$.

\begin{algorithm}[H]\label{algorithm_row_pivot_red}
	\caption{{\small Row pivot reduction}} 
	\DontPrintSemicolon 
	\KwIn{Boundary matrix $\partial$}
	\KwOut{Row reduced boundary matrix $\partial$}
	$R = \partial$ \\
	\For{$i = m,\dots,1$}{ 
		\If{$\leftp{i}\neq 0$}{
			\While{there exists $i' > i$ with $\leftp{i'}=\leftp{i}\neq 0$}{
				add row $i'$ to row $i$}
		} 
	}
\end{algorithm}

In matrix notation, Algorithm \ref{algorithm_row_pivot_red} computes the reduced matrix as $R=W\cdot \partial$, where $W$ is an invertible upper-triangular matrix with $\mathbb{F}_2$-coefficients.
\medskip

For a matrix $D$, consider the following value:
\[
r_{D}\left(i,j\right):= 
\rnk D_{i}^{j} - \rnk D_{i+1}^{j}
+
\rnk D_{i+1}^{j-1} - \rnk D_{i}^{j-1}
\]
where $D_{i}^{j}$ is the lower left submatrix of $D$, given by all the rows of $D$ with index $h\geq i$ and all the columns with index $l \leq j$.
The Pairing Lemma \cite{edels_harer} states that, for a column reduced matrix $C$ of a boundary matrix $\partial$, $i=\lowp{j}$ in $C$ if and only if $r_{\partial}\left(i,j\right)=1$.
An analogous result holds for row reduced matrices (cf.~\cite[Lem 2.2]{Edel-Olsb}):

\begin{lemma}[Row pairing lemma]\label{equivalence_pivot}
	Let $\partial$ be a boundary matrix and $R$ a row reduced matrix of $\partial$.
	Then $j=\leftp{i}$ in $R$ if and only if $r_{\partial}\left(i,j\right)=1$.
\end{lemma}

The proof is precisely the same as the Pairing Lemma \cite{edels_harer} since the used technique relies on the lower-left submatrices.
Moreover, from the Pairing Lemma \cite{edels_harer} and the above Lemma \ref{equivalence_pivot}, $j=\leftp{i}$ in a row reduced matrix $R$ of $\partial$ if and only if $i=low\left(j\right)$ in a column reduced matrix $C$ of $\partial$.
In particular, if $j=\leftp{i}$ or $i=\lowp{j}$, the indices $\left(i, j\right)$ form a {\bf persistence pair}.

\begin{remark}
	Since the coboundary matrix is the anti-transpose of the boundary one (i.e. an element in position $(i,j)$ is sent to position $(m+1-j,m+1-i)$), Algorithm \ref{algorithm_row_pivot_red} performs the standard column reduction on the coboundary matrix. 
	Indeed, for a reduced matrix $R$, $j=\leftp{i}$ if and only if $i=\lowp{j}$ in its anti-transpose $R^{-T}$. 
	Thus, Lemma \ref{equivalence_pivot} provides an alternative proof of the correctness of the standard persistence algorithm in cohomology, result originally showed in \cite{dualities}.
\end{remark}

The running time of Algorithm \ref{algorithm_row_pivot_red} is at most cubic in the number of simplices, as it is for the standard persistence algorithm.
We now refine this estimate a little.

\begin{point}\label{point_computational_costs}
	{\em Computational costs of the reduction.} 
	In \cite{edels_harer}, it is shown that the running time of the inner (i.e. while) loop in the standard persistence algorithm for the column $j$, representing a $h$-simplex $\sigma_{j}$ and whose column pivot is in row $i$, is $(h+1)(j-i)^2$.
	If $\sigma_{j}$ is positive, i.e. at the end of the reduction column $j$ is trivial, then the cost is higher: $(h+1)(j-1)^2$.
	When reducing the coboundary matrix via column operations, as in \cite{ripser_paper, dualities}, the running time becomes $c(j-i)^2$, where $c$ is the number of cofaces of the simplex $\sigma_{j}$, and the cost of reducing a positive column is $c(j-1)^2$.
	When reducing the boundary matrix via row operations, as in Algorithm \ref{algorithm_row_pivot_red}, the running time of the inner loop in Algorithm \ref{algorithm_row_pivot_red} for the row $i$, representing a simplex $\sigma_{i}$ and whose row pivot is in column $j$, is $c(j-i)^2$.
	Note that, if $\sigma_{i}$ is negative, i.e. row $i$ becomes zero at the end of the reduction, then the cost is $c(m-j)^2$, where $m$ is the number of rows.
	Thus, using row operations, the negative rows are the more expensive to reduce, dually to what happens when reducing by columns.
\end{point}

\section{Clear and compress}\label{section_clear_compress}

We now recall two standard runtime optimisations from \cite{clearcompress, twist}, and show their duality using row and column reductions.
Similar observations can be found in \cite{abcw}.
\medskip

A simplex in the filtered simplicial complex $FK$ is called {\bf positive} if it causes the birth of a homological class, and {\bf negative} if it causes the death of a homological class. 
By extension, columns and rows in $\partial$ are called positive (resp. negative) if the corresponding simplices are positive (negative).

\begin{point}\label{point_clearing}
	{\em Clear.}
	The {\bf clear} optimisation is based on the fact that if a row of index $j$ is positive, the $j$-th column of $\partial$ cannot be negative.
	As was already observed in \cite{clearcompress, twist}, this optimisation is particularly effective when performed on the boundary matrices in decreasing degrees, or, as shown in \cite{phat_paper}, when applied to the coboundary matrices in increasing degrees. 
	Since the clear avoids reducing columns that are already known not to contain pivots, it is quite helpful in the persistent algorithms up-to-date.
	However, it is not so useful when reducing by rows, since it shrinks by one the length of each row, but does not avoid any reduction.
\end{point}

\begin{point}\label{point_compress}
	{\em Compress.}
	The {\bf compress} optimisation hinges on the fact that if a column of index $i$ is negative, the $i$-th row of $\partial$ cannot be positive. 
	From \cref{section_eq_relations}, it follows the real advantages of the compress optimisation are obtained when the matrix reduction is performed using row operations. 
	Indeed, in this case, it avoids a costly loop whose results is already known, while, in accordance with previous results \cite{phat_paper}, it is quite inefficient when applied using column operations because it only shortens the columns by one element.
	Performed using the row reduction, the compress is particularly effective when applied to the boundary matrices in increasing degrees. 
\end{point}

Finally, for what we showed, in the computation of the barcode of a filtered simplicial complex the number of rows that need to be reduced in the boundary matrix using the compress optimisation is the same as the number of columns that have to be processed in the coboundary matrix when exploiting the clear optimisation. 
In the case of acyclic complexes, as done in \cite{ripser_paper, abcw}, we can be more precise and show that this number is 
\begin{equation*} 
\displaystyle\sum_{h=0}^{d+1}\binom{v-1}{h}  
\end{equation*}
where $d$ is the maximal dimension of the acyclic filtered simplicial complex and $v$ the number of vertices.
It follows that any algorithm that reduces the coboundary matrix using column operations and the clear can be described as reducing the boundary matrix using row operations and the compress.  

\paragraph*{Acknowledgments.} 
The author was supported by FAR2019-UniMORE and by the Austrian Science Fund (FWF) grant number P 29984-N35.
The author thanks Claudia Landi, Michael Kerber, Wojciech Chach{\'o}lski, and H\aa vard B. Bjerkevik for useful discussions and feedback.

\end{document}